\documentstyle[rotate,epsf]{ichep}
\newcommand{\wh}{\widehat}

\newcommand{\al}{\alpha_{s}}

\newcommand{\ve}{\varepsilon}

\newcommand{\als}{\alpha_{s}(s)}
\newcommand{\aps}{\frac{\als}{\pi}}

\newcommand{\smvs}{\vbox{\vskip 8mm}}

\newcommand{\La}{\Lambda^{(3)}_{\overline{MS}}}

\newcommand{\gsim}{~{}_{\textstyle\sim}^{\textstyle >}~}

\renewcommand{\baselinestretch}{1.2}
\newcommand{\be}{\begin{equation}}
\newcommand{\ee}{\end{equation}}
\newcommand{\ba}{\begin{array}{c}}
\newcommand{\ea}{\end{array}}
\newcommand{\beqn}{\begin{eqnarray}}
\newcommand{\eeqn}{\end{eqnarray}}

\newcommand{\bi}{\begin{itemize}}
\newcommand{\ei}{\end{itemize}}

\newcommand{\cO}{{\cal O}}

\newcommand{\rms}{\rm\scriptsize}

\begin{document}

%%%%
%\vspace*{1.5cm}

%%%%%%%%

\title{\vspace*{-3.2cm}
\begin{flushright}
{\rm\normalsize FTUV/94-50\\
IFIC/94-45\\ October 1994\\}
\end{flushright}
\vspace*{1.0cm}
QCD Analysis of Inclusive $\Delta S=1,2$ Transitions:
  The $|\Delta I|=1/2$ Rule*}

\author{A Pich}

\affil{Departament de F\'{\i}sica Te\`orica and IFIC,
Universitat de Val\`encia -- CSIC, \\
Dr. Moliner 50, E-46100 Burjassot, Val\`encia, Spain}

\abstract{The interplay of QCD in $\Delta S=1,2$ non-leptonic
weak transitions can be rigorously analyzed, at the inclusive
level, by studying the 2--point functions associated
with the corresponding $\Delta S=1,2$ effective Hamiltonians.
The next-to-leading order calculation of these correlators shows
a huge ($\gsim 100\%$) gluonic enhancement of the $|\Delta I|=1/2$
channel, providing a qualitative understanding
of the $|\Delta I|=1/2$ rule within QCD.}

\twocolumn[\maketitle]

\fnm{7}{Invited talk given at ICHEP '94, Glasgow, July 1994.}

\section{Introduction}

The origin of the empirically observed enhancement of
strangeness-changing non-leptonic weak amplitudes with isospin
transfer $|\Delta I|=1/2$ is a long-standing question
in particle physics.
The short-distance analysis of the product of weak hadronic currents
results in an effective $\Delta S=1$ Hamiltonian
\be\label{eq:hamiltonian}
{\cal H}^{\Delta S=1}_{\mbox{\rms eff}} \;
= \; {G_F\over\sqrt{2}}\,V_{ud}^{\phantom{*}} V_{us}^{*} \sum_{i}
C_{i}(\mu^2) \, Q_{i} \,,
\ee
which is a sum of local four-quark operators $Q_i$,
constructed with the light ($u,d,s$) quark fields only,
modulated by Wilson coefficients $C_i(\mu^2)$ which are functions
of the heavy ($t,Z,W,b,c$) masses and an overall renormalization
scale $\mu$.

In the absence of strong interactions, $C_2(\mu^2)=1$ and all other
Wilson coefficients vanish.
The operator $Q_2$ can be decomposed as
$Q_2 = (Q_+ + Q_-)/2$, where
$Q_-\equiv Q_2-Q_1$ is a pure $|\Delta I|=1/2$ operator and
$Q_+\equiv Q_2+Q_1$ induces both $|\Delta I|=1/2$ and $|\Delta I|=3/2$
transitions.
The standard electroweak model gives then rise to $|\Delta I|=1/2$
and $|\Delta I|=3/2$ amplitudes of nearly equal size,
while experimentally the ratio between both amplitudes is a
factor of twenty.
To solve this big discrepancy, QCD effects should be enormous.

The leading $\alpha_s$ corrections indeed give, for $\mu$-values
around 1 GeV, an enhancement by a factor two to three of the
$Q_-$ Wilson coefficient with respect to the $Q_+$ one.
Moreover, the gluonic exchanges generate the additional
$|\Delta I|=1/2$ operators $Q_i$ (i=3,4,5,6), the so-called
``Penguins''.
Nevertheless, this by itself is not enough to explain the
experimentally observed rates, without simultaneously appealing to
a further enhancement in the hadronic matrix elements of at least
some of the isospin--$1/2$ four-quark operators.

The evaluation of hadronic matrix elements is unfortunately
very difficult, since it involves
non-perturbative dynamics at low energies.
The problem gets, moreover, complicated by the $\mu$-dependence
of the matrix elements, which should exactly cancel the
corresponding renormalization-scale dependence of the
Wilson coefficients.
In order to get meaningful results, a full QCD calculation
is required; this is a highly non-trivial task.

The problem becomes much easier at the inclusive level, where the
properties of
${\cal H}^{\Delta S=1}_{\mbox{\rms eff}}$
%the non-leptonic effective weak Hamiltonian
can be analyzed through the 2--point function
\be
%\begin{eqnarray}
\label{eq:correlator}
\eqalign{
%\Psi^{\Delta S=1}(q^2)  \equiv i \int \! dx \, e^{iqx} \,
\Psi(q^2)  &\equiv i \int \! dx \, e^{iqx} \,
\big<0\vert \, T\{\,{\cal H}^{\Delta S=1}_{\mbox{\rms eff}}(x)\,
{\cal H}^{\Delta S=1}_{\mbox{\rms eff}}(0)^\dagger\}\vert0\big>
\cr
&=  \left({G_F\over\sqrt{2}}\right)^2
\left| V_{ud}^{\phantom{*}} V_{us}^{*}\right|^2\,
\sum_{i,j} \, C_i(\mu^2) \, C_j^*(\mu^2) \,\Psi_{ij}(q^2) \,.
\cr}
\ee
%\end{eqnarray}
%
This vacuum-to-vacuum correlator can be studied with
perturbative QCD methods, allowing for a consistent combination of
Wilson coefficients $C_i(\mu^2)$ and 2--point functions of the
4--quark operators, $\Psi_{ij}$, in such a way that the
renormalization scheme and scale dependences exactly cancel (to the
computed order). The associated spectral function,
\be
\frac{1}{\pi}\mbox{\rm Im}\Psi(q^2)
= (2\pi)^3 \sum_\Gamma \int\, d\Gamma
\left|\langle 0| {\cal H}^{\Delta S=1}_{\mbox{\rms eff}}|
\Gamma\rangle\right|^2 \delta^4(q-p_\Gamma) ,
\ee
is a quantity with
definite physical information; it describes in an inclusive way how
the weak Hamiltonian couples the vacuum to physical states
$\Gamma$ of a given
invariant mass. General properties like the observed enhancement of
$|\Delta I|=1/2$ transitions can be then rigorously analyzed at the
inclusive level.

A detailed analysis of  two-point functions
associated with $\Delta S=1$ and $\Delta S=2$ operators was
presented in Ref.~\cite{PD:91},
where the $\cO(\alpha_s)$ corrections to the
correlators $\Psi_{ij}$ were calculated.
The next-to-leading order (NLO) corrections to the
$|\Delta I|=1/2$ 2--point functions
were found to be very large, confirming the QCD enhancement
obtained in a previous approximate calculation \cite{PI:89}.
Those results were, however, incomplete because the NLO
corrections to the Wilson-coefficients of ``Penguin'' operators
were still not known.

The recent calculation of
${\cal H}^{\Delta S=1}_{\mbox{\rms eff}}$
at NLO \cite{BJLW,CFMR}
has allowed us to improve the results of Ref.~\cite{PD:91},
matching matrix elements and Wilson coefficients
consistently at NLO \cite{JP:94}.
Previously missing contributions from evanescent
operators have been also incorporated \cite{JP:94}.
In order to have a check of the results,
the calculation has been performed in two different
renormalization schemes for $\gamma_5$
(naively anticommuting $\gamma_5$ and 't Hooft--Veltman),
and the scale- and scheme-independence of the final
physical quantities has been verified.

\section{Approximate results}

The full calculation of $\Psi(q^2)$ is rather involved due to the
fact that there are several operators which mix under renormalization.
One needs to compute, at the four-loop level, all possible
2--point functions $\Psi_{ij}$; i.e. a $6\times 6$
($12\times 12$ at intermediate steps to include the contributions
of evanescent operators)
matrix correlator which must be renormalized in matrix form,
and later convoluted with the NLO Wilson coefficients
as indicated in Eq.~\ref{eq:correlator}.

It is possible to obtain some simplified results by using
two different approximations which eliminate the mixing
among operators, while keeping at the same time the important
physical effects \cite{PI:89}:

i) If ``Penguins'' are neglected, the operators $Q_\pm$ are
multiplicatively renormalizable.
The corresponding scheme- and scale-independent
spectral functions
$\Phi_{\pm\pm}(s) \equiv C_\pm^2(\mu^2)
{1\over \pi}\mbox{\rm Im}\Psi_{\pm\pm}(s,\mu^2)$
are found to be \cite{JP:94}:
\begin{eqnarray}
\Phi_{++}(s) \sim
\frac{8}{15}\,\frac{s^4}{(4\pi)^6}\,\alpha_s(s)^{-4/9}\,
\biggl[\,1-\frac{3649}{1620}\,\aps\,\biggr] \,,
&\label{eq:phi_pp} \\
\smvs
\Phi_{--}(s)  \sim
\frac{4}{15}\,\frac{s^4}{(4\pi)^6}\,\alpha_s(s)^{8/9\phantom{-}}\,
\biggl[\,1+\frac{9139}{810}\,\aps\,\biggr] \,.
&\label{eq:phi_mm}
\end{eqnarray}

ii) The interesting ``Penguin'' operator $Q_6$ can be isolated, by
noting that in the large $N_c$ limit ($N_c$ = number of colours)
the anomalous dimension matrix $\gamma_{ij}$ of the set of
operators $Q_i$ becomes zero, but for $\gamma_{66}$; i.e.
in this limit
there is no mixing among operators  and only $Q_6$ gets
renormalized.
%(Note that
%the large $N_c$ limit is not appropriate for studying the $Q_\pm$
%operators  since $\gamma_{\pm\pm}=0$ in this limit).
The $\cO(\alpha_s^2)$ correction can also be
easily computed in this limit \cite{PD:91}:
\be\label{eq:Phi_penguin}
\eqalign{
\fl\Phi_{66}(s)\sim {3\over 5} {s^4\over (4\pi)^6}
\alpha_s(s)^{18/11}
&\biggl[ 1 + \frac{117501}{4840}\aps \cr
&+ 470.72 \left(\aps\right)^2 \biggr] .\cr }
\ee

The NLO corrections to the $|\Delta I|=1/2$ correlators turn out to
be very big and positive, while for $\Phi_{++}$ the
correction is moderate and negative.
Taking $\alpha_s(s)/\pi\approx 0.1$, we find a moderate suppression
of $\Phi_{++}$ by roughly 20\%, whereas $\Phi_{--}$
acquires a huge enhancement of the order of 100\%.
The correction is even bigger for the ``Penguin'' correlator $\Phi_{66}$:
240\% at NLO and 700\% at next-to-next-to-leading order!
The perturbative calculation blows up in the
$|\Delta I|=1/2$ sector, clearly showing a dynamical gluonic
enhancement of the $|\Delta I|=1/2$ amplitudes.

\section{Exact results}

Following the notation of Refs.~\cite{BJLW}, the
Wilson-coefficient functions can be
decomposed as $C_i(s) = z_i(s) + \tau\,y_i(s)$, where
$\tau\equiv - \left(V_{td}^{\phantom{*}} V_{ts}^*\right)/
\left(V_{ud}^{\phantom{*}} V_{us}^*\right)$. The coefficients
$z_i(s)$ govern the real part of the effective Hamiltonian,
while $y_i(s)$
parametrize the imaginary part and govern, e.g., the measure for
direct CP-violation in the $K$-system, $\ve'/\ve$.
We can then form two
different scale-
and scheme-invariant spectral functions,
\beqn\label{eq:Phi_z}
\wh\Phi_z(s) & = & \sum_{i,j}\;
z_i(\mu^2)\,{1\over\pi}\mbox{\rm Im}\Psi_{ij}(s,\mu^2)\,z_j(\mu^2) \, ,
 \\ \label{eq:Phi_y}
\wh\Phi_y(s) & = & \sum_{i,j}\;
y_i(\mu^2)\,{1\over\pi}\mbox{\rm Im}\Psi_{ij}(s,\mu^2)\,y_j(\mu^2)\, ,
\eeqn
corresponding to $z_i$ and $y_i$ respectively.

Since we are mainly interested in the size of the
radiative corrections, let us write
$\wh\Phi_{z,\,y}(s)$ as
\begin{equation}
\wh\Phi_{z,\,y}(s)\;=\;
\wh\Phi^{(0)}_{z,\,y}(s)+\wh\Phi^{(1)}_{z,\,y}(s)
\,,
\label{eq:Phi_z_y}
\end{equation}
where the superscripts $(0)$ and $(1)$ refer to the
leading and next-to-leading order respectively.
The exact results obtained \cite{JP:94} for the ratios
$\wh\Phi^{(1)}_{z}/\wh\Phi^{(0)}_{z}$ and
$\wh\Phi^{(1)}_{y}/\wh\Phi^{(0)}_{y}$
are plotted in Fig.~1, for $\La=200,\,300$, and $400$
MeV.

%%%%%%%%%%%%%%%%%%  FIGURE  %%%%%%%%%%%%%%%%%%%%%%%%%%%%%
\begin{figure}[bht]
\centerline{
\rotate[r]{
\epsfysize=8.5cm
\epsffile{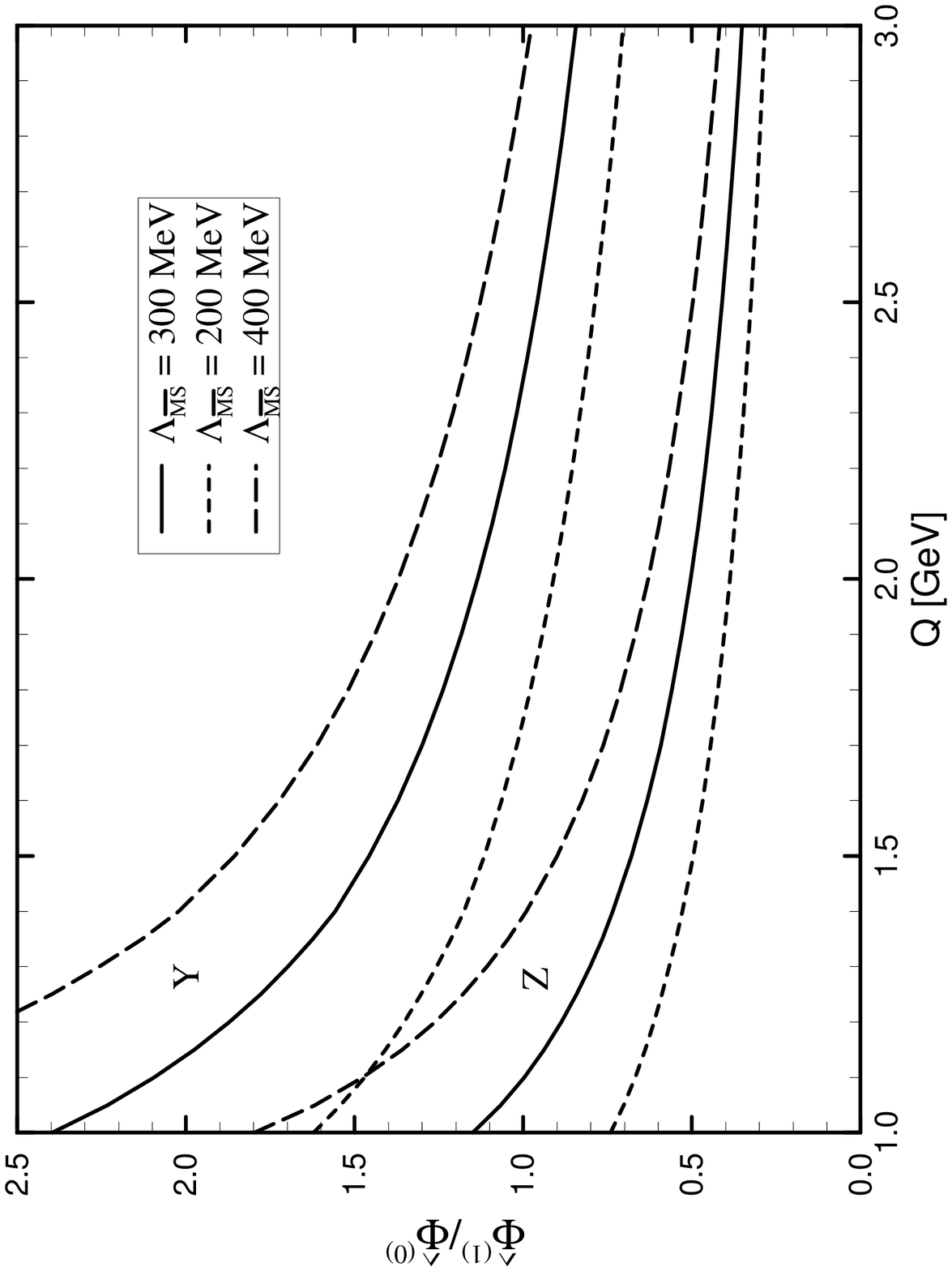}
}}
\caption[]{The ratios $\wh\Phi^{(1)}_{z}/\wh\Phi^{(0)}_{z}$ and
$\wh\Phi^{(1)}_{y}/\wh\Phi^{(0)}_{y}$.\label{figure}}
\end{figure}

%%%%%%%%%%%%%%%%% END FIGURE  %%%%%%%%%%%%%%%%%%%%%%%%%

{}From Fig.~1, we can see that in the region $Q=1-3\,\mbox{\rm GeV}$,
and for a central value $\La=300\,\mbox{\rm MeV}$, the radiative QCD
correction to $\wh\Phi_z$ ranges approximately between 40\% and 120\%,
whereas in the case of $\wh\Phi_y$ we find a correction of the order
of 100\%--240\%.
As explicitly shown by the approximate results of the previous
section,
the large $\al$ corrections
correspond to the $|\Delta I|=1/2$ part of the effective weak
Hamiltonian.
In fact, the corrections to the $|\Delta I|=3/2$
correlator are exactly given by Eq.~\ref{eq:phi_pp}
(``Penguins'' only give $\Delta I=1/2$ contributions), and
therefore are quite moderate.

In the case of $\Delta S=2$ transitions, there is only one 4--quark
operator.
% in the effective Hamiltonian.
Since the $\Delta S=2$ and $|\Delta I|=3/2$
operators belong to the same representation  of the
(flavour) $SU(3)_L\otimes SU(3)_R$ group, the NLO corrections
to the $\Delta S=2$ correlator are also exactly given by
Eq.~\ref{eq:phi_pp}.

\section{Summary}

The short-distance behaviour of the $\Delta S=1$ correlators clearly
shows a dynamical enhancement of the $|\Delta I|=1/2$
channel, as a consequence of the interplay of gluonic corrections.
The structure of the radiative corrections also allows for a deeper
understanding of the underlying dynamical mechanism \cite{JP:94}:
large corrections appear wherever quark-quark correlations can contribute.
This explains why the phenomenological description
of the $|\Delta I|=1/2$ rule
in terms of intermediate
effective diquarks \cite{NS:91} was so successful.

A full QCD calculation has been possible because of the inclusive
character of the defined 2--point functions. Although only qualitative
conclusions can be directly extracted from these results, they are
certainly important since they rigorously point to the
QCD origin of the infamous $|\Delta I|=1/2$ rule.
%and, moreover, provide valuable information on the relative importance
%of the different operators, which can be very helpful to
%attempt more pragmatic calculations.

\section*{Acknowledgments}

I would like to thank M. Jamin for a very enjoyable collaboration.
This work has been supported in part by  CICYT (Spain) under
Grant No. AEN-93-0234.

%%%%%%%%%%%%%% References %%%%%%%%%%%%%%%%%

\end{document}